\documentclass[prb,preprint]{revtex4-1} 

\usepackage{float}

\usepackage{ amssymb }
\usepackage{upgreek}

\usepackage{amsmath}  
\usepackage{amsfonts} 
\usepackage{graphicx} 

\begin{document}


\title{Using Simulations to Highlight the Role of Electromagnetism in Special Relativity}




\date{\today}
\author{Refath Bari}
\begin{abstract}
 Physics Education Research (PER) studies have demonstrated that undergraduate students struggle with fundamental principles from Electromagnetism (EM) and Special Relativity (SR). However, few studies have approached the intersection of the two subjects: the role which EM played in inspiring the development of SR. To address this issue, this paper presents two simulations which highlight the fundamental role of electromagnetism in special relativity. The first simulation demonstrates the Lorentz Transformations and the origin of the Gamma Factor from Maxwell's Equations. The second simulation offers an experiential introduction to the Biot-Savart Law, from which the Displacement Current in Ampere-Maxwell's Law can be derived. These programs may be useful in an undergraduate electromagnetism course. The simulations discussed in this paper are available at the link given in the footnote\footnote{https://refath.notion.site/Using-Simulations-to-Highlight-the-Role-of-Electromagnetism-in-Special-Relativity-107a01c7bd394fdab5b6e7bbfe158fe4?pvs=4}.
\end{abstract}

\maketitle 

\section{Introduction} 
A significant program in Physics Education Research (PER) has been research on student difficulties with undergraduate electromagnetism \citep{BestEM, PollockMan, Hernandez2023, Chabra2023}. Furthermore, the counter-intuitive features of Special Relativity have also been well documented \citep{Paolo2014, Parizot2010, Gousopoulos2015}. However, the subtle insight which inspired Einstein to make the leap from Classical Electrodynamics to Special Relativity\cite{Dunstan2008-ju}, namely the frame dependence of electric and magnetic fields, is not emphasized in most undergraduate electromagnetism textbooks \citep{Tanel2014}. As a result, many students struggle to understand how relativistic phenomenon such as time dilation and the Lorentz gamma factor emerge from classic electromagnetism.

Undergraduate textbooks seldom highlight the historical connection between Special Relativity and Electromagnetism \cite{Jackson, Purcell, Griffiths, Fleisch}. Many studies have demonstrated the effectiveness of simulations in clarifying student confusions of abstract concepts such as space-time intervals and causality \citep{Alstein2023, Savage2007, Big2020}. Although a few emerging texts and individual papers are paving the way for experiential-based electromagnetism via computational visualizations\cite{ChabaySherwood, Xuan2023, Morris2012, Franklin2019}, most textbooks have yet to take advantage of simulations to better highlight the intimate relationship between the two subjects. To address this issue, this paper presents two simulations which highlight the role of electromagnetism in special relativity, supplemented with the prerequisite theory for completeness. The two simulations are as follows: (1) A simulation of the Lorentz Transformation in 1+1 Dimensions and (2) A simulation of the Biot-Savart Law. These simulations may be useful in an undergraduate course on Electromagnetism. 

The above two principles were selected due to previously documented student difficulties associated with them. For instance, the Lorentz transformation is a subject of considerable consternation amongst students \cite{Boudreaux2002, Castells2010, Tanel}. Its formulation as a symmetric matrix transformation may seem to be a mere accident\cite{Scherr2002}. Furthermore, it may seem to be an unmotivated ansatz, spuriously satisfying lorentz invariance\cite{Seluk2010}. The application of the lorentz transformation to transition between reference frames is another well-documented student difficulty \cite{Sherin2016, Weiskopf2006}. To address this issue, we present a simulation of the Lorentz transformations in 1+1 dimensions. We also present a theoretical motivation for the Lorentz transformation, showing why anything less than a full transformation of both space and time (i.e., Galilean Transformation and Fitzgerald Transformation), fails to preserve the invariance of the wave equation. 
\section{Motivating the Lorentz Transformation}
The motivating basis for the Lorentz Transformations is that the Galilean Transformations fails to maintain the invariance of Maxwell's Equations. Many students grapple with the inconsistency of the Galilean Transformations, but to no avail. Indeed, many struggle with the basic Galilean Principle of Relativity itself\cite{Parizot2010}. To address this problem, we demonstrate why transforming neither space or time (Galilean Transformation) or transforming only space (Fitzgerald Transformation) fails to preserve the invariance of the wave equation under a transformation. The wave equation is trivial to obtain from Maxwell's Equations: 
\begin{align}
    \begin{array}{l}
\nabla \times B=\frac{\partial E}{\partial t}+{\mu}_{0}J \rightarrow -\nabla(\nabla \cdot E)+\nabla^{2} E=E_{t t}
\end{array}
\end{align}        
We now demonstrate why the Galilean and Fitzgerald Transforms fail to preserve this wave equation, and thus motivate why a transformation of both the space and time is required.
\subsection{Application of GT to Wave Equation}
To simplify our analysis, we only consider a wave dependent on $(z,t)$.
\begin{align}
\frac { \partial ^{ 2 }E }{ \partial ^{ 2 }t } =\nabla ^{ 2 }E \rightarrow { E }_{ tt }-{ E }_{ zz }=0
\end{align}
The Galilean Transformation has $\zeta = z+vt,\tau=t$
\begin{align}
    E(t,z)=\varepsilon (t,z+vt)=\varepsilon (\tau ,\zeta )
\end{align}
We use the multivariable chain rule to verify whether ${ E }_{ tt }-{ E }_{ zz }=0$. We begin with
\begin{align}
    { E }_{ t }=\frac { \partial \varepsilon  }{ \partial \tau  } \cdot \frac { \partial \tau  }{ \partial t } +\frac { \partial \varepsilon  }{ \partial \zeta  } \cdot \frac { \partial \zeta  }{ \partial t } ={ \varepsilon  }_{ \tau  }+{ \varepsilon  }_{ \zeta  }\cdot \frac { \partial (z-vt) }{ \partial t } ={ \varepsilon  }_{ \tau  }-v{ \varepsilon  }_{ \zeta  } \rightarrow { E }_{ tt }=\frac { \partial  }{ \partial t } ({ \varepsilon  }_{ \tau  }-v{ \varepsilon  }_{ \zeta  })
\end{align}
Finding ${E}_{tt}$ requires evaluating $\frac { \partial  }{ \partial t } ({ \varepsilon  }_{ \tau  })$ and $\frac { \partial  }{ \partial t } (-v{ \varepsilon  }_{ \zeta  })$. We find that ${E}_{tt}$ becomes
\begin{align}
    { E }_{ tt }=\frac { \partial  }{ \partial t } ({ \varepsilon  }_{ \tau  })+\frac { \partial  }{ \partial t } (-v{ \varepsilon  }_{ \zeta  })=({ \varepsilon  }_{ \tau \tau  }-v{ \varepsilon  }_{ \tau \zeta  })-(-v{ \varepsilon  }_{ \zeta \tau  }+{ v }^{ 2 }{ \varepsilon  }_{ \zeta \zeta  })={ \varepsilon  }_{ \tau \tau  }-v{ \varepsilon  }_{ \tau \zeta  }+v{ \varepsilon  }_{ \zeta \tau  }-{ v }^{ 2 }{ \varepsilon  }_{ \zeta \zeta  }
\end{align}
Likewise, we find ${E}_{zz}$. Note that $\uptau = t = 0$ and that $\uptau$ has no dependence on $z$, so that ${\uptau}_{z}=0$ and ${ \mathcal{E}  }_{ \uptau  }{ \uptau  }_{ z }=0$. By the Galilean Transformation, $\zeta = z - vt$, and thus we have 
\begin{align}
    { E }_{ z }=\frac { \partial \varepsilon  }{ \partial \tau  } \cdot \frac { \partial \tau  }{ \partial z } +\frac { \partial \varepsilon  }{ \partial \zeta  } \cdot \frac { \partial \zeta  }{ \partial z }={ \varepsilon  }_{ \tau  }\cdot 0+{ \varepsilon  }_{ \zeta  }\cdot \frac { \partial (z-vt) }{ \partial z } ={ \varepsilon  }_{ \zeta  }
\end{align}
\begin{align}
    { E }_{ zz }=\frac { \partial ^{ 2 }E }{ \partial z^{ 2 } } =\frac { \partial  }{ \partial t } ({ \varepsilon  }_{ \zeta  })=\frac { \partial ({ \varepsilon  }_{ \zeta  }) }{ \partial \tau  } \cdot \frac { \partial \tau  }{ \partial z } +\frac { \partial ({ \varepsilon  }_{ \zeta  }) }{ \partial \zeta  } \cdot \frac { \partial \zeta  }{ \partial z }={ \varepsilon  }_{ \zeta \tau  }\cdot 0+{ \varepsilon  }_{ \zeta \zeta  }\cdot \frac { \partial (z-vt) }{ \partial z } ={ \varepsilon  }_{ \zeta \zeta  }
\end{align}
Thus, the Galilean Transformation fails to preserve the Wave Equation:
\begin{align}
    { E }_{ tt }-{ E }_{ zz }={ \varepsilon  }_{ \tau \tau  }-v{ \varepsilon  }_{ \tau \zeta  }+v{ \varepsilon  }_{ \zeta \tau  }-{ v }^{ 2 }{ \varepsilon  }_{ \zeta \zeta  }-{ \varepsilon  }_{ \zeta \zeta  }\neq 0
\end{align}
The next significant transformation came from George Francis FitzGerald, who made conjectured that length itself may contract. We will now implement the Fitzgerald Transformation and find that it also fails to preserve the wave equation.

\subsection{Application of FT to Wave Equation}
The Fitzgerald Transformation (FT) was coined as an ad-hoc correction \cite{Michelson-Morley} to the Galilean Transformation by George Francis Fitzgerald in his 1889 paper on the Ether and the Earth's Atmosphere\cite{Ether,Fitzgerald}. Fitzgerald published a colloquial paper, stating that length contracted "by an amount depending on the square of the ratio of their velocities to that of light":
\begin{align}
E(t,z)=\varepsilon (t,\frac { z }{ \sqrt { 1-{ v }^{ 2 } }  } -\frac { vt }{ \sqrt { 1-{ v }^{ 2 } }  } ) = \varepsilon(\tau,\zeta)
\end{align}
To verify whether the Fitzgerald Transformations holds the Wave Equation invariant, we must verify whether ${E}_{tt}-{E}_{zz}=0$ by expanding ${E}_{t}$ using the multivariable chain rule. 
\begin{align}
    { E }_{ t }={ \varepsilon  }_{ \tau  }+{ \varepsilon  }_{ \zeta  }\cdot \frac { \partial (\frac { z }{ \sqrt { 1-{ v }^{ 2 } }  } -\frac { vt }{ \sqrt { 1-{ v }^{ 2 } }  } ) }{ \partial t } ={ \varepsilon  }_{ \tau  }+{ \varepsilon  }_{ \zeta  }\cdot (-\frac { v }{ \sqrt { 1-{ v }^{ 2 } }  } )\rightarrow { E }_{ t }={ \varepsilon  }_{ \tau  }-\frac { v{ \varepsilon  }_{ \zeta  } }{ \sqrt { 1-{ v }^{ 2 } }  } 
\end{align}
\begin{align}
    { E }_{ tt }=\frac { \partial ^{ 2 }E }{ \partial t^{ 2 } } =\frac { \partial  }{ \partial t } ({ \varepsilon  }_{ \tau  }-\frac { v{ \varepsilon  }_{ \zeta  } }{ \sqrt { 1-{ v }^{ 2 } }  } )={ \varepsilon  }_{ \tau \tau  }-\frac { 2v{ \varepsilon  }_{ \tau \zeta  } }{ \sqrt { 1-{ v }^{ 2 } }  } +\frac { v^{ 2 }{ \varepsilon  }_{ \zeta \zeta  } }{ 1-{ v }^{ 2 } } 
\end{align}
Likewise, we find ${E}_{zz}$ using the chain rule to be
\begin{align}
{ E }_{ z }={ \varepsilon  }_{ \tau  }\cdot 0+{ \varepsilon  }_{ \zeta  }\cdot \frac { \partial (\frac { z }{ \sqrt { 1-{ v }^{ 2 } }  } -\frac { vt }{ \sqrt { 1-{ v }^{ 2 } }  } ) }{ \partial z } =\frac { { \varepsilon  }_{ \zeta  } }{ \sqrt { 1-{ v }^{ 2 } }  } \rightarrow { E }_{ zz }=\frac { { \varepsilon  }_{ \zeta \zeta  } }{ 1-{ v }^{ 2 } } 
\end{align}
We now verify if FT preserves the invariance of the Wave Equation:
\begin{align}
{ E }_{ tt }-{ E }_{ zz }=({ \varepsilon  }_{ \tau \tau  }-\frac { 2v{ \varepsilon  }_{ \tau \zeta  } }{ \sqrt { 1-{ v }^{ 2 } }  } +\frac { v^{ 2 }{ \varepsilon  }_{ \zeta \zeta  } }{ 1-{ v }^{ 2 } } )-(\frac { { \varepsilon  }_{ \zeta \zeta  } }{ 1-{ v }^{ 2 } } ) ={ \varepsilon  }_{ \tau \tau  }-{ \varepsilon  }_{ \zeta \zeta  }-\frac { 2v{ \varepsilon  }_{ \tau \zeta  } }{ \sqrt { 1-{ v }^{ 2 } }  } \neq 0
\end{align}
We find that even length contraction has failed to uphold the invariance of the wave equation. However, it is instructive to examine the matrix formulation of $F (R^2\rightarrow R^2)$:
\begin{align}
F:{ R }^{ 2 }\rightarrow { R }^{ 2 }=\begin{pmatrix} 1 & 0 \\ \frac { -v }{ \sqrt { 1-{ v }^{ 2 } }  }  & \frac { 1 }{ \sqrt { 1-{ v }^{ 2 } }  }  \end{pmatrix}\begin{pmatrix} t \\ z \end{pmatrix}=\begin{pmatrix} t \\ \frac { z-vt }{ \sqrt { 1-{ v }^{ 2 } }  }  \end{pmatrix}=\begin{pmatrix} \tau  \\ \zeta  \end{pmatrix}
\end{align}
Hendrik Lorentz independently made the next conjecture, that both space and time must be transformed to preserve $E_{tt}=E_{zz}$. We observe that the matrix formulations for the FT and LT are quite similar. Students will thus be able to understand the origin of the Lorentz Transformation, not as an ansatz, but as an evolution from GT and FT.
\begin{align}
L:{ R }^{ 2 }\rightarrow { R }^{ 2 }:\begin{pmatrix} \frac { 1 }{ \sqrt { 1-{ v }^{ 2 } }  }  & \frac { -v }{ \sqrt { 1-{ v }^{ 2 } }  }  \\ \frac { -v }{ \sqrt { 1-{ v }^{ 2 } }  }  & \frac { 1 }{ \sqrt { 1-{ v }^{ 2 } }  }  \end{pmatrix}
\end{align}
Immediately, we find the transformation to be symmetric such that $L={L}^{T}$ and $L={L}^{-1}$. But does it maintain the invariance of the wave equation? To verify, we briefly apply LT to the Wave Equation. 

\subsection{Application of LT to Wave Equation}
We now consider the Lorentz Transformations (LT) and test whether $E_{tt}=E_{zz}$ under $L$.
\begin{align}
L:{ R }^{ 2 }\rightarrow { R }^{ 2 }:\begin{pmatrix} \frac { 1 }{ \sqrt { 1-{ v }^{ 2 } }  }  & \frac { -v }{ \sqrt { 1-{ v }^{ 2 } }  }  \\ \frac { -v }{ \sqrt { 1-{ v }^{ 2 } }  }  & \frac { 1 }{ \sqrt { 1-{ v }^{ 2 } }  }  \end{pmatrix}\begin{pmatrix} t \\ z \end{pmatrix}=\begin{pmatrix} \frac { t-vz }{ \sqrt { 1-{ v }^{ 2 } }  }  \\ \frac { z-vt }{ \sqrt { 1-{ v }^{ 2 } }  }  \end{pmatrix}\end{align}

\begin{align}
E(t,z)=\varepsilon (\tau ,\zeta )=\varepsilon (\frac { t-vz }{ \sqrt { 1-{ v }^{ 2 } }  } ,\frac { z-vt }{ \sqrt { 1-{ v }^{ 2 } }  } ) \rightarrow { E }_{ t }=\frac { \partial \varepsilon  }{ \partial \tau  } \cdot \frac { \partial \tau  }{ \partial t } +\frac { \partial \varepsilon  }{ \partial \zeta  } \cdot \frac { \partial \zeta  }{ \partial t } 
\end{align}
However, the two observers' times are no longer equivalent. By the Lorentz Transformation, we now have time dilated as one's velocity $v$ approaches the speed of light $c$.
\begin{align}
{ E }_{ t }={ \varepsilon  }_{ \tau  }\cdot \frac { \partial (\frac { t }{ \sqrt { 1-{ v }^{ 2 } }  } -\frac { vz }{ \sqrt { 1-{ v }^{ 2 } }  } ) }{ \partial t } +{ \varepsilon  }_{ \zeta  }\cdot \frac { \partial (\frac { z }{ \sqrt { 1-{ v }^{ 2 } }  } -\frac { vt }{ \sqrt { 1-{ v }^{ 2 } }  } ) }{ \partial t }=\frac { { \varepsilon  }_{ \tau  }-{ \varepsilon  }_{ \zeta  }v }{ \sqrt { 1-{ v }^{ 2 } }  } 
\end{align}
Taking the second partial derivative of the Electric Field in respect to time, we have
\begin{align}
{ E }_{ tt }=\frac { \partial  }{ \partial \tau  } ({ E }_{ t })=\frac { \partial  }{ \partial \tau  } (\frac { { \varepsilon  }_{ \tau  }-{ \varepsilon  }_{ \zeta  }v }{ \sqrt { 1-{ v }^{ 2 } }  } )=\frac { { \varepsilon  }_{ \tau \tau  }-{ \varepsilon  }_{ \zeta \tau  }v }{ 1-{ v }^{ 2 } } +\frac { -v{ \varepsilon  }_{ \tau \zeta  }+{ \varepsilon  }_{ \zeta \zeta  }v^{ 2 } }{ 1-{ v }^{ 2 } } =\frac { { \varepsilon  }_{ \tau \tau  }-2{ \varepsilon  }_{ \zeta \tau  }v+{ \varepsilon  }_{ \zeta \zeta  }v^{ 2 } }{ 1-{ v }^{ 2 } } 
\end{align}
Likewise, we find ${E}_{zz}$ using the chain rule.
\begin{align}
{ E }_{ z }=\frac { \partial \varepsilon  }{ \partial \tau  } \cdot \frac { \partial \tau  }{ \partial z } +\frac { \partial \varepsilon  }{ \partial \zeta  } \cdot \frac { \partial \zeta  }{ \partial z } ={ \varepsilon  }_{ \tau  }\cdot -\frac { v }{ \sqrt { 1-{ v }^{ 2 } }  } +{ \varepsilon  }_{ \zeta  }\cdot \frac { 1 }{ \sqrt { 1-{ v }^{ 2 } }  } =\frac { { \varepsilon  }_{ \zeta  }-v{ \varepsilon  }_{ \tau  } }{ \sqrt { 1-{ v }^{ 2 } }  } 
\end{align}
\begin{align}
{ E }_{ zz }=\frac { \partial  }{ \partial z } (\frac { { \varepsilon  }_{ \zeta  }-v{ \varepsilon  }_{ \tau  } }{ \sqrt { 1-{ v }^{ 2 } }  } )=\frac { \partial (\frac { { \varepsilon  }_{ \zeta  }-v{ \varepsilon  }_{ \tau  } }{ \sqrt { 1-{ v }^{ 2 } }  } ) }{ \partial \tau  } \cdot \frac { \partial \tau  }{ \partial z } +\frac { \partial (\frac { { \varepsilon  }_{ \zeta  }-v{ \varepsilon  }_{ \tau  } }{ \sqrt { 1-{ v }^{ 2 } }  } ) }{ \partial \zeta  } \cdot \frac { \partial \zeta  }{ \partial z }=\frac { -2v{ \varepsilon  }_{ \zeta \tau  }+v^{ 2 }{ \varepsilon  }_{ \tau \tau  }+{ \varepsilon  }_{ \zeta \zeta  } }{ 1-{ v }^{ 2 } } 
\end{align}
Completing the proof, we find that 
\begin{align}
{ E }_{ tt }-{ E }_{ zz }=\frac { { \varepsilon  }_{ \tau \tau  }-2{ \varepsilon  }_{ \zeta \tau  }v+{ \varepsilon  }_{ \zeta \zeta  }v^{ 2 } }{ 1-{ v }^{ 2 } } -\frac { -2v{ \varepsilon  }_{ \zeta \tau  }+v^{ 2 }{ \varepsilon  }_{ \tau \tau  }+{ \varepsilon  }_{ \zeta \zeta  } }{ 1-{ v }^{ 2 } } =0
\end{align}
We have just proved the Wave Equation does indeed hold invariant under the Lorentz Transformation $L$. From this transformation alone, rises all major tenets of Special Relativity, including Time Dilation and Length Contraction. 

\subsection{Simulation of Lorentz Transformation in 1+1 Dimensions}
The simulation below demonstrates the lorentz transformation in 1+1 dimensions. Students may use the simulation to transform any event $E:(x,t)$ between reference frames. Three sliders are presented at the bottom of the simulation, which students may use to control the location of an event in space-time and the relative velocity between two observers. Students will subsequently observe an animation of the coordinate axes of the simulation bending to illustrate the new coordinates. 
\begin{figure}[h]
    \centering
    \includegraphics[width=1\textwidth]{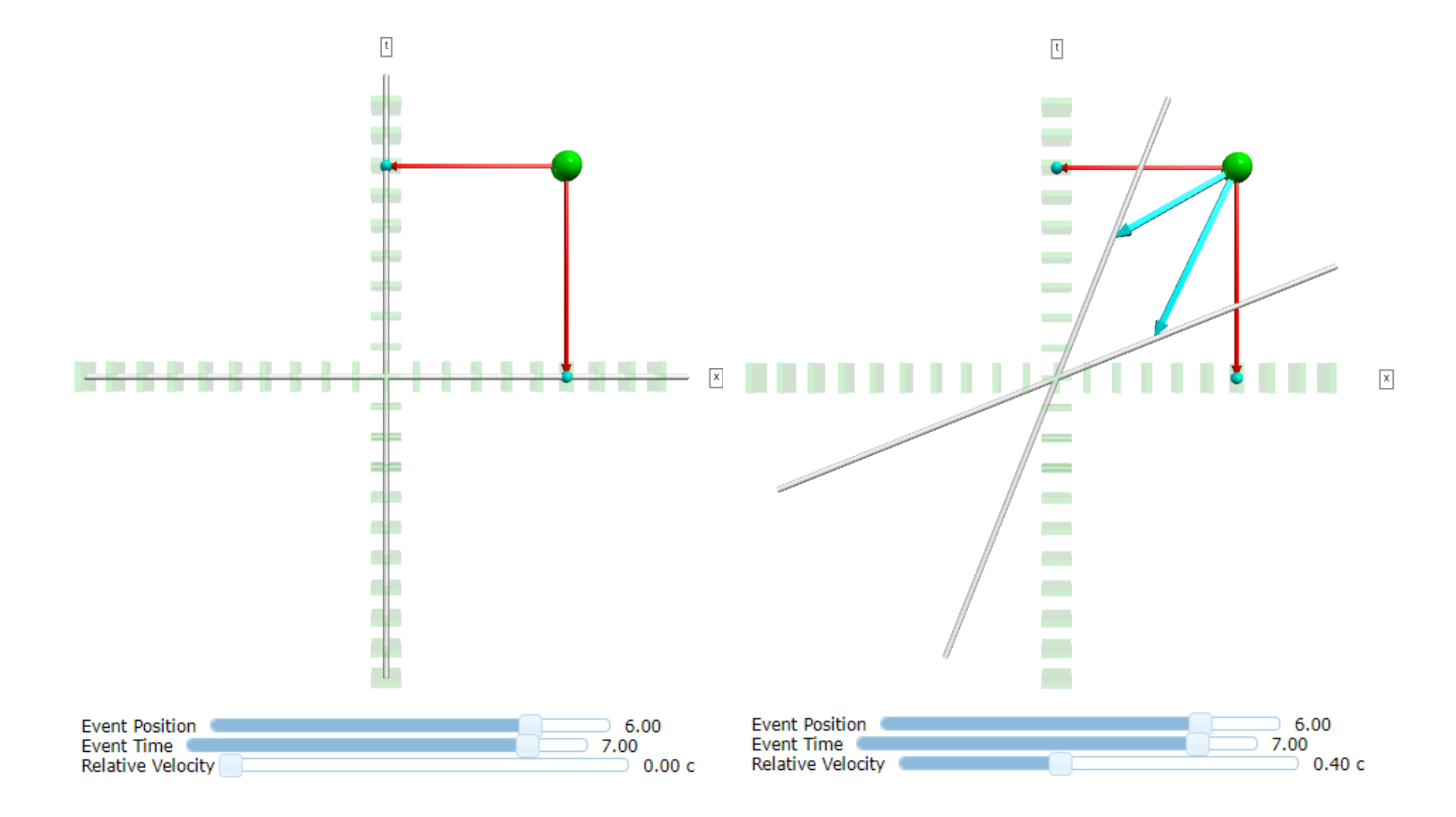}
    \caption{Lorentz Transformation Simulation in 1+1 Dimensions}
    \label{fig:enter-label}
\end{figure}

\section{$\frac{\partial \vec{E}}{\partial{t}}$ from Biot-Savart and STR}
The traditional introduction of the Displacement Current term in standard electromagnetics textbooks is via an inconsistency in Ampere's Law when applied to Parallel Plate Capacitors \cite{Buschauer2014} (an Amperian Loop can be chosen such that a current does not penetrate the boundary of the surface). This presentation may lead one to believe the Displacement Current to be an ad-hoc correction to Ampere's Law, when it is actually a natural consequence of a moving test charge \cite{Sato1980}. 

To motivate the existence of the displacement current, we chose an alternative route -- not a parallel plate capacitor, but a simple test charge located at the origin.

According to Maxwell himself, the Displacement Current term is 'electrostatically' analogous to a normal current element or a moving test charge. We thus examine a test charge from a stationary and moving observer's point of view. Whereas a stationary observer only finds an electric field, an observer moving with velocity $-v$ with respect to the stationary frame will witness both an electric and magnetic field. Such a test charge would thus exhibit the displacement current, due to the moving electric field. We take the relativistic form of the electromagnetic fields of the test charge and derive the Biot Savart Law from it. Hidden implicitly within the Biot Savart Law is the Displacement Current Term, which we reformulate using the Partial Derivative of a Cross Product to conclude with Ampere's Law, corrected with the Displacement Current Term. Before doing so, however, it is crucial to understand the Biot-Savart Law as it is traditionally applied to a current-carrying wire.

\subsection{Visualization of Biot-Savart}
We seek to create the magnetic field vector at any distance $R$ from the current-carrying wire. To do so, we require both Magnitude and Direction. The magnitude will be supplied by the Biot-Savart Law as 
\begin{align}
\| \vec{B}\|=\frac{{mu}_{0}I}{2\pi R}
\end{align}
he Right Hand Rule gives the direction of the magnetic field to be counterclockwise, since the current has dir$(I)=\hat{j}$. The unexpected challenge, however, is to compute a consistently counterclockwise magnetic field $\vec{B}$ vector around the wire. To do so, we construct three auxiliary vectors: a vector $\vec{u}$ from the origin to the $y$ axis, a vector $\vec{p}$ from the origin to the cursor position, and a vector $\vec{v}$ as the difference between the two vectors. From that difference, a dot product $\vec{v} \cdot \vec{w}$  is calculated such that the result is $0$, and  thus we have that $\vec{w}$ is indeed the magnetic field vector at the cursor's position.
\begin{figure}[h]
\includegraphics[width=15cm]{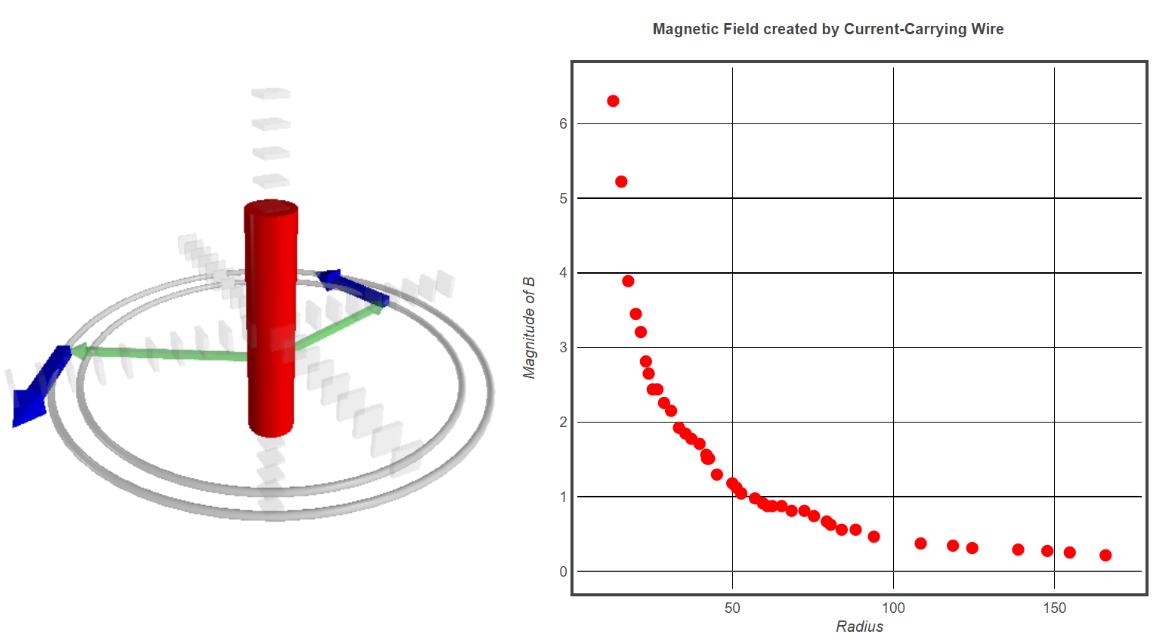}
\centering
\caption{Simulation of Magnitude and Direction of $\vec{B}$ for a Current-Carrying Wire}
\end{figure}
Prior to constructing this in code, we must declare three functions -- each corresponding to three of the user's actions: impressing the track-pad, dragging, and releasing. We thus define our \texttt{down()} function to be 
In the next auxiliary function \texttt{Python}{move()}, the program updates the magnitude and direction of all four vectors based on the position of the cursor. We must also declare a few global variables so they may be used outside the \texttt{Python}{move()} function. 

Finally, we have the release function \texttt{Python}{up()}, which generates a graph of the magnitude of the magnetic field by distance and updates real-time, based on the user's placement of Magnetic Field vectors. We may verify the graph and indeed find that the magnetic field drops proportional to the distance $R$. 
\subsection{Time Dilation from EM}
We now pivot from the Biot-Savart Law to a moving test charge. In doing so, we will find that the Law we have just visually demonstrated implicitly contains within it the Displacement Current, without the need for any capacitors. Consider once again our test charge $Q$, centered at the origin and moving in the $\hat{i}$ direction. As Maxwell himself states the Displacement Current to be "electrostatically" equivalent to a traditional current formed by a moving Electric Field, it makes sense to find the change in the Electric Field of the charge in respect to time. To do so, we employ the chain rule as 
\begin{align}
    \frac { \partial E }{ \partial t } =\frac { \partial E }{ \partial x } \cdot \frac { \partial x }{ \partial t } = \frac { \partial E }{ \partial t } =(-v)\cdot \frac { \partial  }{ \partial x } (\frac { 1 }{ 4\pi { \varepsilon  }_{ 0 } } \frac { Q }{ { r }^{ 2 } } )
\end{align}
Here's where relativity comes in. Through the Electromagnetic Field Strength Tensor, we find the equations of a relativistic electric field to be 
\begin{align}
    {E}_{\parallel}={E'}_{\parallel},{B}_{\parallel}=0, {E}_{\perp}=\gamma{E'}_{\perp}, B=+\gamma \frac{1}{{c}_{2}} v \times E'
\end{align}
We thus have the Magnetic Field at any given distance $R$ to be
\begin{align}
B(r)=\gamma\frac{1}{{c}^{2}}v\times E=\gamma{\mu}_{0} \frac{qv\times \hat{r}}{4\pi {r}^{2}}
\end{align}
We thus find that the Displacement Current is implicit within the Biot-Savart Law.
\begin{align}
    B(r) = \gamma \frac{1}{{c}^{2}} (\vec{v} \times \vec{E}) = \gamma{\mu}_{0}r \times ({\epsilon}_{0} \frac{\partial E}{\partial t})\rightarrow\frac{\partial B}{\partial t}=\gamma{\mu}_{0}v\times ({\epsilon}_{0} \frac{\partial E}{\partial t})
\end{align}
We thus find that the Biot Savart Law implicitly contains the Displacement Current \cite{Neuenschwander1992}.

\section{Conclusion}
We have demonstrated how computational programming visualizations can work symbiotically with theoretical derivations to solidify the concepts behind the bridge between EM and SR and make them tangible -- indeed, computational visualizations such as the ones above offer rich pedagogical opportunities for teachers and students alike to intuitively grasp the ideas of Maxwell's Equations and Einstein's Special Relativity.

\bibliographystyle{apalike}
\bibliography{References}

\end{document}